\documentclass[aps,prl,twocolumn,superscriptaddress,showpacs,preprintnumbers]{revtex4-1}

\usepackage{amsmath,amssymb,ifpdf,url
             ,mathrsfs,slashed,multirow,tabularx,type1cm}
\usepackage{graphicx,color}

\ifpdf        
 \usepackage{graphics}
\else                          
\usepackage[dvipdfmx]{graphics}      
\fi
\allowdisplaybreaks

\definecolor{BlueViolet}{rgb}{0.2, 0.00, 0.7}
\definecolor{Blue}{rgb}{0.15, 0.00, 0.9}
\usepackage[
colorlinks=true,linkcolor=Blue,citecolor=Blue,
urlcolor=BlueViolet]{hyperref}

\newcommand{\TeV}{\,{\rm TeV}}

\newcommand{\Slash}[1]{{\ooalign{\hfil \hspace*{-5pt}~#1\hfil\crcr\raise.167ex\hbox{/}}}}
\def\be{\begin{equation}}
\def\ee{\end{equation}}

\def\({\left(}
\def\){\right)}
\def\<{\langle}
\def\>{\rangle}
\newcommand{\non}{\nonumber \\ }
\newcommand{\matl}{\left( \begin{array}}
\newcommand{\matr}{\end{array} \right)}

\newcommand{\eq}[1]{Eq.~(\ref{#1})}

\def\beq#1\eeq{\begin{align}#1\end{align}}
\newcommand{\gsim}{ \mathop{}_{\textstyle \sim}^{\textstyle >} }

\newcommand{\bra}[1]{ \langle {#1} | }
\newcommand{\ket}[1]{ | {#1} \rangle }

\newcommand{\GEV}{\text{\,GeV} }
\newcommand{\TEV}{\text{\,TeV} }

\newcommand{\kk}{\ensuremath{K^0\textrm{-}\overline{K}{}^0}}
\newcommand{\real}{\textrm{Re}\,}
\newcommand{\imag}{\textrm{Im}\,}
\newcommand{\fig}[1]{Fig.~\ref{#1}}

\begin{document}

\preprint{TTP16--014}

\title{Supersymmetric Explanation of \boldmath$CP$ Violation in 
       $K\to \pi\pi$ Decays\unboldmath}

\author{Teppei Kitahara} \email{teppei.kitahara@kit.edu} 
\affiliation{Institute for Theoretical Particle Physics (TTP), 
Karlsruhe Institute of Technology, Wolfgang-Gaede-Stra\ss e 1, 
 76128 Karlsruhe, Germany}
\affiliation{Institute for Nuclear Physics (IKP), Karlsruhe Institute of
  Technology, Hermann-von-Helmholtz-Platz 1, 76344
  Eggenstein-Leopoldshafen, Germany}

\author{Ulrich Nierste}
\email{Ulrich.Nierste@kit.edu}
\affiliation{Institute for Theoretical Particle Physics (TTP), 
Karlsruhe Institute of Technology, Wolfgang-Gaede-Stra\ss e 1, 
 76128 Karlsruhe, Germany}

\author{Paul Tremper} \email{paul.tremper@kit.edu}
\affiliation{Institute for Theoretical Particle Physics (TTP),
  Karlsruhe Institute of Technology, Wolfgang-Gaede-Stra\ss e 1, 76128
  Karlsruhe, Germany}

\date{\today}

\begin{abstract}
  {Recent progress in the determination of hadronic matrix elements
    has revealed a tension between the measured value of
    $\epsilon_K^\prime/\epsilon_K$, which quantifies direct
    $CP$ violation in $K \to \pi\pi$ decays, and the Standard-Model
    prediction. The well-understood indirect $CP$ violation encoded in the
    quantity $\epsilon_K$ typically precludes large new-physics
    contributions to $\epsilon_K^\prime/\epsilon_K$ and challenges such
    an explanation of the discrepancy. We show that it is possible to
    cure the $\epsilon_K^\prime/\epsilon_K$ anomaly in the Minimal
    Supersymmetric Standard Model with squark masses above 3$\,\TEV$
    without overshooting $\epsilon_K$. This solution exploits two
    features of supersymmetry: the possibility of large isospin-breaking
    contributions (enhancing $\epsilon_K^\prime$) and the Majorana
    nature of gluinos (permitting a suppression of $\epsilon_K$). Our
    solution involves no fine-tuning of $CP$ phases or other parameters.}
\end{abstract}


\maketitle

\vspace{-0.2cm}
\section{Introduction}
\vspace{-0.2cm}
Measurements of charge-parity ($CP$) violation are sensitive probes of
physics beyond the Standard Model (SM).  $CP$ violation in $K\to \pi\pi$
decays is characterized by the two quantities, $\epsilon_K$ and
$\epsilon_K^\prime$, which describe indirect and direct $CP$ violation,
respectively.  $| \epsilon_K|=(2.228\pm 0.011)\times 10^{-3}$ measures $CP$
violation in the \kk\ mixing amplitude, in which the strangeness quantum
number $S$ changes by two units \cite{Agashe:2014kda}.  $\epsilon_K^\prime$ quantifies $CP$
violation in the $|\Delta S|=1$ amplitude triggering the decay $K\to
\pi\pi$. To predict $\epsilon_K^\prime$ in the SM one must calculate
hadronic matrix elements of four-quark operators with nonperturbative
methods. A determination of all operators by lattice QCD has  been
obtained only recently \cite{Bai:2015nea}, and the predicted
$\epsilon_K^\prime$ lies substantially below the experimental
value \cite{epspexp}: %
\beq%
\frac{\epsilon_K^\prime}{\epsilon_K}=
\begin{cases}
  \left(16.6 \pm 2.3 \right) \times 10^{-4}&(\textrm{PDG\,\cite{Agashe:2014kda}})\\
 {\left(1.0 \pm 4.7 \pm1.5 \pm   0.6  \right)}\times 10^{-4}&(\textrm{SM-NLO})
  \end{cases}
  \label{discrepancy}
  \eeq %
  Our SM prediction {\cite{SMepsilonprime,footnote:uncertainty}} is
  based on the next-to-leading order (NLO) calculation of Wilson
  coefficients and anomalous dimensions \cite{nlo, Buras:1993dy} and
  {the hadronic matrix elements of Refs.~\cite{Bai:2015nea,
      Blum:2011ng}. {As in Ref.~\cite{Buras:2015yba}, we} exploit
    $CP$-conserving data to reduce hadronic uncertainties.  The two
    numbers in \eq{discrepancy} disagree by $2.9\,\sigma$
    {\cite{SMepsilonprime,Buras:2015yba}}.  This tension is
    underpinned by results found with the $1/N_c$ expansion {(dual
      QCD approach)} \cite{bbg,bgnew,bgnew2}, which is a completely different
    calculational method \cite{Buras:2015yba}.}  In the near future the
  increasing precision of lattice calculations will sharpen the SM
  prediction in \eq{discrepancy} further and answer the question about
  new physics (NP) in $\epsilon_K^\prime$.

  An explanation of the puzzle in \eq{discrepancy} by physics beyond the
  SM calls for a NP contribution which is seemingly even larger than the
  SM value. On general grounds, however, one expects that NP effects in
  a $|\Delta F|=1$ four-quark process are highly suppressed once
  constraints from the corresponding $|\Delta F|=2$ transition are taken
  into account. Here $F$ denotes the flavor quantum number, and $F=S$ in
  our case of $K\to \pi\pi$ decays. To explain the NP hierarchy in
  $|\Delta F|=1$ vs $|\Delta F|=2$ transitions, we specify to
  $\epsilon_K^\prime$ and $\epsilon_K$: The SM contributions to both
  quantities are governed by the combination%
  \beq%
  \tau = - \frac{V_{td}V_{ts}^*}{V_{ud}V_{us}^*} \sim (1.5 - i 0.6)\times
  10^{-3} \eeq%
  of elements of the Cabibbo-Kobayashi-Maskawa (CKM) matrix with
  $\epsilon_K^{\prime\,\rm SM} \propto \imag \tau/M_W^2$ and
  $\epsilon_K^{\rm SM} \propto \imag \tau^2/M_W^2$.  If the NP
  contribution comes with the $\Delta S=1$ parameter $\delta$ and is
  mediated by heavy particles of mass $M$, one finds
  $\epsilon_K^{\prime\, \mathrm{NP}} \propto \imag \delta/M^2$,
  $\epsilon_K^{\mathrm{NP}} \propto \imag \delta^2/M^2$ and therefore
  {the experimental constraint} {$|\epsilon_K^{\mathrm{NP}}| \leq
    |\epsilon_K^{\rm SM}|$ leads to}%
\beq%
  {\left| {\frac{\epsilon^{\prime\,
          \mathrm{NP}}_K}{\epsilon^{\prime\, \mathrm{SM}}_K}
      } \right|} \leq  
  \frac{\left|\epsilon_K^{\prime\,
         \mathrm{NP}}/\epsilon_K^{\prime\,\rm SM} \right|}{
     \left|\epsilon_K^{\mathrm{NP}}/\epsilon_K^{\rm SM} \right|}
  = {\cal O}\left( \frac{\real \tau}{\real \delta} \right).
\label{eq:sens}
\eeq%
With $M\gsim 1\TEV$, NP effects can be relevant only for $|\delta|\gg
|\tau|$, and \eq{eq:sens} seemingly forbids detectable NP contributions
to $\epsilon_K^\prime$. In this Letter, we show that \eq{eq:sens} can be
overcome in the Minimal Supersymmetric Standard Model (MSSM) and {one
  can} reproduce the central value of the measured $\epsilon_K^\prime$
in \eq{discrepancy} with squark and gluino masses in the multi-TeV
range.  {Our solution involves no fine-tuning of $CP$ phases or other
  parameters.}

\vspace{-0.4cm}
\boldmath  
\section{$\epsilon_K^\prime$ in the 
              MSSM}\unboldmath
\vspace{-0.2cm}
The MSSM is a good candidate for physics beyond the SM, because it 
alleviates the hierarchy problem, improves gauge coupling unification,
and provides dark-matter candidates. Present collider bounds
\cite{Khachatryan:2015vra} (and the largish Higgs mass of 125\GEV
\cite{Aad:2012tfa, Chatrchyan:2012xdj}) push the masses of colored
superpartners into the TeV range, which makes supersymmetry an imperfect
solution to the hierarchy problem but actually improves gauge coupling
unification.

The master equation for $\epsilon_K^\prime$ reads \cite{Buras:2015yba}
\beq%
\frac{\epsilon_K^\prime}{\epsilon_K} = \frac{\omega_{+}}{\sqrt{2}
  {|}\epsilon_K^{\textrm{exp}}{|} \textrm{Re} A_0^{\textrm{exp}} } \left\{
  \frac{\textrm{Im} A_2 }{\omega_{+}} - \left( 1-
    \hat{\Omega}_{\textrm{eff}} \right) \textrm{Im} A_0 \right\},
\label{eq:mas}
\eeq%
with $\omega_{+}= (4.53 \pm 0.02)\times10^{-2}$, {the measured
  $|\epsilon_K^{\rm exp}|$,} $\hat{\Omega}_{\textrm{eff}} = (14.8\pm
8.0)\times 10^{-2}$, and the amplitudes $A_I = \langle (\pi \pi)_I |
\mathcal{H}^{\left|\Delta S\right| = 1} | K^0 \rangle$ involving the
effective $|\Delta S|=1$ {Hamiltonian} $ \mathcal{H}^{\left|\Delta
    S\right|}$. $I=0,2$ labels the strong isospin of the final two-pion
state. $\textrm{Im} A_2$ is under good control for some time
\cite{Blum:2011ng}; the recent theory progress of
Refs.~\cite{Bai:2015nea,bgnew,bgnew2} concerns the QCD penguin contribution
{to} $\textrm{Im} A_0$.  {Prior to the first reliable lattice result
  for $\textrm{Im} A_0$ \cite{Bai:2015nea}, SM predictions for
  $\epsilon_K^\prime$ were based on analytic methods, the dual QCD
  method of Refs.~\cite{bbg,bgnew,bgnew2}, or chiral perturbation theory
  \cite{Pallante:1999qf}. The second method gives a larger value for
  $\epsilon_K^\prime$ because of an enhancement of $\textrm{Im} A_0$
  from final-state interaction. In the calculation of
  Ref.~\cite{Pallante:1999qf}, this effect is strictly correlated with a
  (phenomenologically welcome) enhancement of $\textrm{Re} A_0$. In the
  dual QCD method, this correlation is absent \cite{bgnew2}. With
  shrinking errors, lattice gauge theory  will settle the issue of $\textrm{Im} A_0$ soon. It is
  important to state that the lattice calculation of Ref.~\cite{Bai:2015nea} 
  \emph{does}\ include final-state interaction along the line of
  Ref.~\cite{Lellouch:2000pv}.}

The MSSM contribution to $\epsilon_K^\prime$ simply adds to the SM
piece.
{Supersymmetric} contributions to $\epsilon_K^\prime/\epsilon_K$ have
been widely studied \cite{Gabbiani:1996hi,Buras:1998ed,Masiero:1999ub,
  Buras:1999da, Kagan:1999iq,Grossman:1999av} in the past, but for a
supersymmetry-breaking scale $M_S$ in the ballpark of the electroweak
scale, so that the suppression mechanism inferred from \eq{eq:sens} is
avoided. 

\begin{figure}[tb]
\vspace{-0.4cm}
  \begin{center} 
    \includegraphics[width=0.40 \textwidth, 
    bb 
    = 0 0 1753 544]{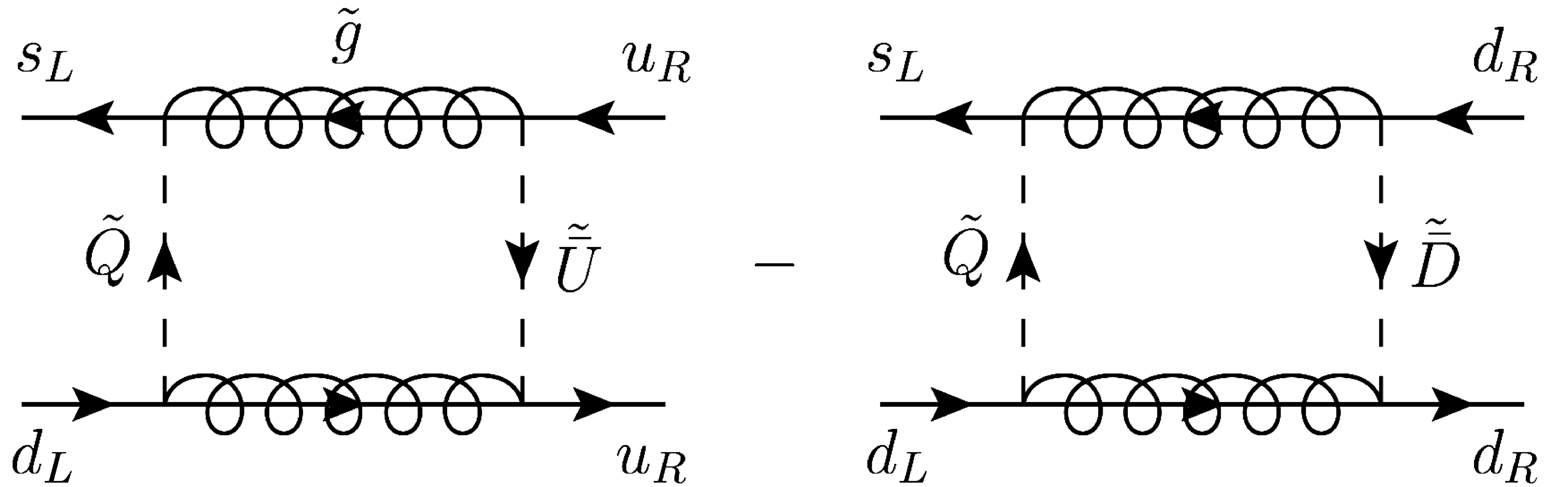}
    \caption{Supersymmetric {gluino} box  contribution to
      $\epsilon_K^\prime/\epsilon_K$ (called a {\it Trojan
        penguin}\ in Ref.~\cite{Grossman:1999av}). It contributes to $\imag A_2$ for
      $m_{\bar{U}}\neq m_{\bar{D}}$ and is the largest contribution in
      our scenario. 
      There are also crossed box diagrams.
      \label{fig:trojan}}
\end{center}
\vspace{-0.7cm}
\end{figure}

\begin{figure*}[t]
  \begin{center} 
   \includegraphics[width=0.45 \textwidth, 
   bb
    = 0 0 254 166]{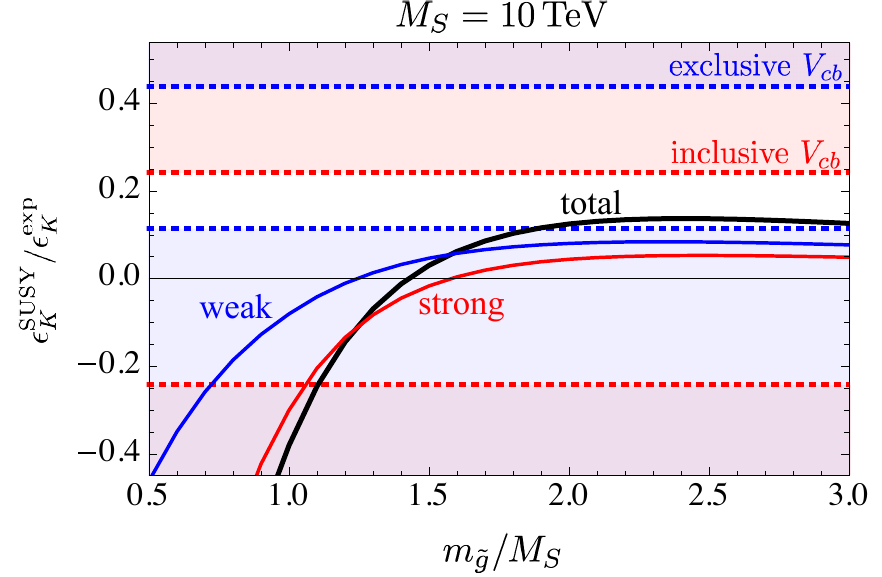}
   \hspace{1.cm}
    \includegraphics[width=0.45 \textwidth, 
    bb
     = 0 0 251 157]{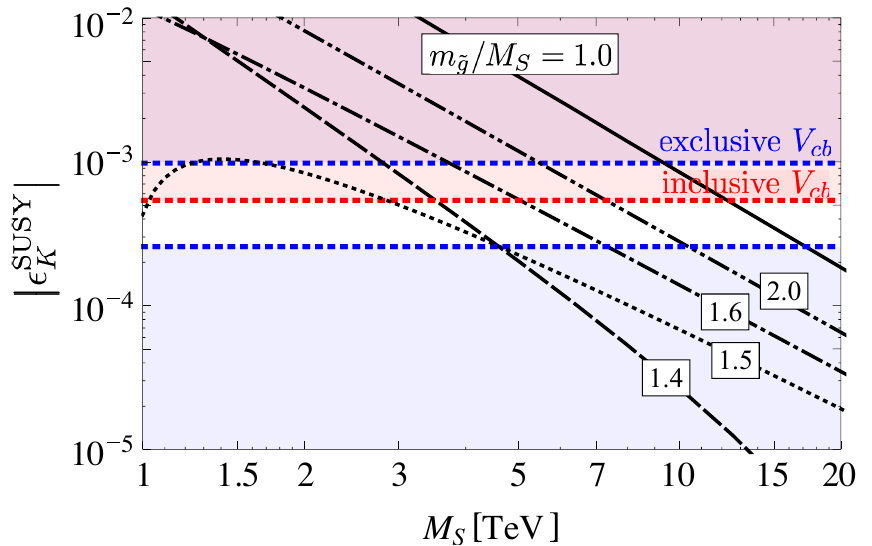}
    \vspace{-0.2cm}
    \caption{The {\it left} plot shows
      $\epsilon^{\textrm{SUSY}}_K/\epsilon^{\textrm{exp}}_K$ as a
      function of the gluino-squark mass ratio $m_{\tilde{g}}/M_S$,
      where we take $M_S = m_{Q} = m_{\bar{U}} = m_{\bar{D}} =
      10\TeV$. The red {line shows the
        gluino-gluino box contribution (with the zero crossing
        near $m_{\tilde{g}}/M_S=1.5$ \cite{Crivellin:2010ys}), 
        while the blue line denotes the sum
        of the box contributions with one or two winos. The total
        contribution is shown in black.}  The red (blue) regions are
      excluded by the measurement of $\epsilon_K$ at the 95\,\% 
      confidence level (C.L.),
      {if the SM prediction uses} the inclusive (exclusive)
      {measurement of} $|V_{cb}|$~\cite{Bailey:2015tba}.  On the {\it
        right}, the black lines {show} $\left|
        \epsilon_K^{\textrm{SUSY}}\right|$ {for several}
      gluino-squark mass ratios as a function of the squark mass.
\label{fig:epsilonK}}
\end{center}
\vspace{-0.7cm}
\end{figure*}

In the absence of sizable left-right squark mixing the low-energy
  Hamiltonian reads
\beq
\mathcal{H}_{\textrm{eff, SUSY}}^{\left|\Delta S \right| = 1} =
\frac{G_F}{\sqrt{2}}\sum_q \left[ \sum_{i=1}^{2} c_i^{q} (\mu) Q_i^q
  (\mu) \right. ~~~~~~~~~~~~~~~& \non \left.  +\sum_{i=1}^4  [
    c_i^{\prime q} (\mu) Q_i^{\prime q}(\mu) + \tilde{c}_i^{\prime q}
    (\mu) \tilde{Q}_i^{\prime q}(\mu) ] \right] + \textrm{H.c.},
\label{eq:hsusy}
\eeq
where $G_F$ is the Fermi constant and
\begin{eqnarray}
&Q^{ q}_1 = \left( \bar{s}_{\alpha} q_{\beta} \right)_{{}_{V-A}}\left( \bar{q}_{\beta} d_{\alpha} \right)_{{}_{V-A}},~
Q^{ q}_2 = \left( \bar{s}  q \right)_{{}_{V-A}} \left( \bar{q}  d \right)_{{}_{V-A}},\non
&Q^{\prime q}_1 = \left( \bar{s}  d \right)_{{}_{V-A}} \left( \bar{q}  q \right)_{{}_{V+A}},~
Q^{\prime q}_2 = \left( \bar{s}_{\alpha}  d_{\beta} \right)_{{}_{V-A}} \left( \bar{q}_{\beta}  q_{\alpha} \right)_{{}_{V+A}}, \non
&Q^{\prime q}_3 = \left( \bar{s}  d \right)_{{}_{V-A}} \left( \bar{q}  q \right)_{{}_{V-A}},~
Q^{\prime q}_4 = \left( \bar{s}_{\alpha}  d_{\beta} \right)_{{}_{V-A}}  \left( \bar{q}_{\beta}  q_{\alpha} \right)_{{}_{V-A}}. \non
\end{eqnarray}
Here $( \bar{s} d)_{V-A} ( \bar{q} q )_{V\pm A}=
  [\bar{s}\gamma_{\mu}(1-\gamma_5)d][ \bar{q} \gamma^{\mu}(1\pm \gamma_5) q]$,
  $\alpha$ and $\beta$ are color indices, and opposite-chirality
  operators $ \tilde{Q}_i^{\prime q}$ are found by interchanging $V-A
  \leftrightarrow V+A$. {In the presence of moderate
    left-right mixing, also the chromomagnetic penguin
    operator $Q_{8g}=m_sg_s/(16 \pi^2) \bar s T^a
    \sigma_{\mu\nu}(1-\gamma_5)d G^{\mu\nu\, a}$ can be
    relevant and is included in our discussion below.}
Our solution exploits two special features of
supersymmetric theories.  First, there are loops governed by the strong
interaction which contribute to $\textrm{Im} A_2$ entering \eq{eq:mas}
with the enhancement factor $1/\omega_+ = 22.1$
\cite{Kagan:1999iq,Grossman:1999av}.  These are gluino-box diagrams
which feed the $(\pi\pi)_{I=2}$ final state if the right-handed up and
down squarks ($\tilde{\bar{U}}$ and $\tilde{\bar{D}}$) have different masses (see
\fig{fig:trojan}). The flavor-changing neutral-current parameter is the $(1,2)$ element of the
left-handed down squark mass matrix $M^2_{Q} $ inducing $\tilde
s_L$-$\tilde d_L$ mixing.  Second, the Majorana nature of the gluino
leads to a suppression of the gluino-squark contribution to
$\epsilon_K$, because there are \emph{two}\ such diagrams (crossed and
uncrossed boxes) with opposite signs. If the gluino mass $m_{\tilde g}$
equals roughly 1.5 times the average down squark 
mass $M_S$ and if either left-handed or right-handed squark mixing 
  is suppressed, both contributions to $\epsilon_K^{\rm SUSY}$ cancel
\cite{Crivellin:2010ys}. For $m_{\tilde g}>1.5 M_S $, the gluino-box
contribution approximately behaves as $[m_{\tilde g}^2-(1.5
M_S)^2]/m_{\tilde g}^4$, with a shallow maximum at $m_{\tilde
  g}{\simeq}2.5 M_S $, after which the $1/m_{\tilde g}^2$ decoupling
sets in.  In this parameter region also chargino, neutralino, and
gluino-neutralino box diagrams are important \cite{Crivellin:2010ys} and
are included in our numerics.  The up-type squark mass matrix is $(V
M^2_{Q} V^{\dag})_{ij}$ [up to negligible ${\cal O} (v^2)$ terms, where
$v$ is the electroweak vacuum expectation value], so that also chargino diagrams are affected
by squark flavor mixing.  The measured $\epsilon_K$ agrees well with
the SM expectation, if the global CKM fit uses the $|V_{cb}|$ measured
in inclusive semileptonic $B$ decays \cite{Alberti:2014yda}, but exceeds $\epsilon_K^{\rm SM}$
for the smaller $|V_{cb}|$ inferred from exclusive decays
\cite{Bailey:2015tba, Charles:2015gya}. Figure~\ref{fig:epsilonK} shows that for both cases
$\epsilon_K^{\rm SM}+\epsilon_K^{\rm SUSY}$ complies with
$\epsilon_K^{\rm exp}$ over a wide parameter range without fine-tuning.

\begin{figure}[tb]
  \begin{center} 
    \includegraphics[width=0.45 \textwidth, 
    bb
     = 0 0 360 233]{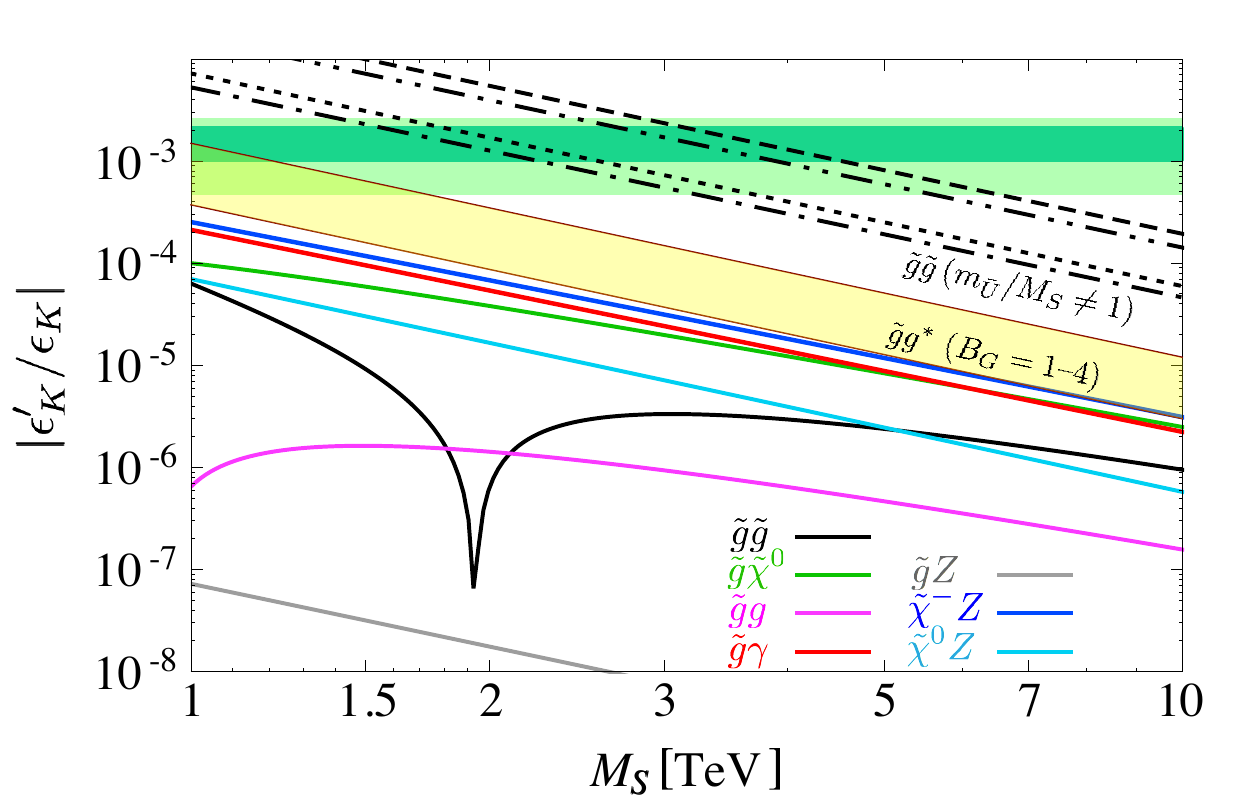}
    \vspace{-0.2cm}
    \caption{Individual supersymmetric contributions to $\left|
        \epsilon_K^\prime/\epsilon_K\right|$ as a function of
      {$M_S=m_Q=m_{\bar D}$}.  
      $\tilde{g}\tilde{g}$, $\tilde{g}\tilde{\chi}^0$,
      $\tilde{g}g$, $\tilde{g}\gamma$, $\tilde{g}Z$,
      $\tilde{\chi}^{-}Z$, $\tilde{\chi}^0 Z$, and $\tilde{g} g^{\ast}$
      represent the gluino-gluino and gluino-neutralino boxes, gluino
      gluon, photon, and $Z$ penguins, chargino and neutralino
      $Z$ penguins, and chromomagnetic contributions, {respectively}.  
      The thick lines {show} 
      the case of universal squark masses, $m_{\bar{U}} = M_S $.
      The broken black lines are the gluino-gluino box {contributions for}
      $m_{\bar{U}} /M_S = 0.5,\,2.0,\,0.8,\,1.2 $ from top to bottom.
      The  $\epsilon_K^\prime/\epsilon_K$ discrepancy is resolved at
      $1\,\sigma$\,($2\,\sigma$) in the dark (light) green band. }
\label{fig:each_contribution}
\end{center}
\vspace{-0.7cm}
\end{figure}

{To get the desired large effect in $\epsilon_K^\prime$ we need a
  contribution to the operators $Q_{1,2}^\prime$ with $(V-A)\times(V+A)$
  Dirac structure, whose matrix elements are chirally enhanced by a
  factor $(m_K/m_s)^2$. Therefore, the flavor mixing has to be in the
  left-handed squark mass matrix. The opposite situation with
  right-handed flavor mixing and $\tilde u_L$-$\tilde d_L$ mass
  splitting is not possible, because SU(2)$_L$ invariance enforces
  $M_{\tilde u_L}^2 - M_{\tilde d_L}^2={\cal O} (v^2)$.  Therefore,
    our scenario involves flavor mixing between left-handed squarks
    only.  We use the following notation for the squark mass matrices:
  $ M^2_{X, ij} = m^2_{X} \left( \delta_{ij} + \Delta_{X, ij} \right), $
  with $X = Q,~\bar{U}$, or $\bar{D}$. {Throughout this Letter we use
    $m^2_{Q} =m^2_{\bar D}=M_S^2$ and vary $m_{\bar U}$}.  We have
  calculated all one-loop contributions to the coefficients in
  \eq{eq:hsusy} in the squark mass eigenbasis and will present the
  full results elsewhere \cite{preparation2}. 
  For the dominant ``Trojan penguin'' contribution,
  we confirm the result of Ref.~\cite{Kagan:1999iq} and find a typo in
  the expression for $c_4^\prime$ in Ref.~\cite{Grossman:1999av}.  The
  second-largest contribution to $\epsilon_K^\prime$ stems from the
  chromomagnetic penguin operator, and our coefficient is in agreement
  with Refs.~\cite{Gerard:1984ky,Abel:1998iu}. To our knowledge, the
  other coefficients have been obtained only in the mass insertion
  approximation \cite{Gabbiani:1996hi}, and our results agree upon
  expansion in $\Delta_{X, ij}$. {Our results also comply with the loop
    diagram results collected in Ref.~\cite{goto}.}  The individual
  contributions {to $\epsilon_K^\prime/\epsilon_K$} are shown in
  \fig{fig:each_contribution}. }

{For the calculation of $\epsilon_K^\prime/ \epsilon_K$, we must use the
  renormalization group (RG) equations to evolve the Wilson coefficients
  calculated at the high scale $\mu=M_S$ down to the hadronic scale
  $\mu_h = {\cal O} (1\,\GEV)$ at which the operator matrix elements are
  calculated.}  In order to {use} the well-known {NLO}
$10\times10$ anomalous dimensions for the SM four-fermion operator basis
\cite{nlo} {we switch from \eq{eq:hsusy} to}%
\beq%
\mathcal{H}_{\textrm{eff, SUSY}}^{\left| \Delta S \right| =1 } =
\frac{G_F}{\sqrt{2}} \sum_{i=1}^{10} [ C_{i} (\mu) Q_i (\mu) +
  \tilde{C}_{i} (\mu) \tilde{Q}_i (\mu) ] + \textrm{H.c.}, %
\eeq%
where $Q_{1,\dots,10}$ are given in Refs.~\cite{nlo, Buras:1993dy} 
and 
\beq
C_{1,2} (\mu) &= c_{1,2}^u (\mu),~~~~\tilde{C}_{1,2}(\mu)= 0, \non
C_{3,4,5,6} (\mu) &= \frac{1}{3} [ c_{3,4,1,2}^{' u}(\mu) + 2
  c_{3,4,1,2}^{' d} (\mu) ],\non C_{7,8,9,10} (\mu) &=
\frac{2}{3}[ c_{1,2,3,4}^{'u}(\mu) - c_{1,2,3,4}^{'d} (\mu)
], && 
\eeq
and {the coefficients} $\tilde{C}_{3,\dots10}$ {of the
  opposite-chirality operators} are {found from $C_{3,\dots10}$ by
  replacing} $c^{'q}_i \to \tilde{c}^{'q}_i$.  Note that $C_{7,8} $
receive the contribution of Fig.~\ref{fig:trojan}.

{For the evolution of the coefficients from $\mu=M_S$ to $\mu=\mu_h$, we
  use a new analytical solution  of the RG equations which avoids the
  problem of a singularity in the NLO terms discussed in
{Refs.~\cite{SMepsilonprime,Adams:2007tk}}. For $ \mathcal{H}_{\textrm{eff,
      SUSY}}^{\left| \Delta S \right| =1 }$, we employ proper threshold
  matching at the scales $\mu_{t,b,c}$ set by the top, bottom, and charm
  quark masses with the usual threshold matching matrices
  \cite{Buras:1993dy}.  In our analysis we take $\mu_h = 1.3$ GeV. For
  the SM prediction in \eq{discrepancy} and the calculation of the MSSM
  prediction, we have evolved the matrix elements of
  Refs.~\cite{Bai:2015nea, Blum:2011ng} (which are given at
    $\mu=1.531\,\GEV$ for $A_0$ and at $\mu=3.0\,\GEV$ for $A_2$) to
  $\mu_h$ with three-flavor full NLO operator mixing. The use of
  NLO RG formulae for $ \mathcal{H}_{\textrm{eff, SUSY}}^{\left| \Delta
      S \right| =1 }$ involves a relative error of order 
  $\alpha_s(M_S)$, because the two-loop corrections to the
  initial conditions of the Wilson coefficients are not
  included. However, the NLO corrections proportional to the much larger
  $\alpha_s(\mu_h)$  are all correctly
  included and independent of the renormalization scheme.}

\vspace{-0.2cm}
\boldmath  
\section{Phenomenology of  $\epsilon_K$ 
               and $\epsilon_K^\prime$}
\unboldmath  
{In this section, we study $\epsilon_K$ and
  $\epsilon_K^\prime/\epsilon_K$ in the MSSM parameter region in which
  the discrepancy in \eq{discrepancy} is removed.
  As input, we} take $\alpha_s \left( M_Z \right) = 0.1185$, the grand-unified theory 
relation for gaugino masses, $m_{\tilde{g}}/M_S = 1.5 $, and $m_{Q} =
m_{\bar{D}} = \mu_{\rm SUSY}=M_S$, {where $\mu_{\rm SUSY}$ is the
  Higgsino mass parameter. Furthermore, the trilinear
  supersymmetry-breaking matrices $A_q$ are set to zero, $\tan \beta
  =10$, and the only nonzero off-diagonal elements of the squark mass
  matrices are $\Delta_{Q, 12,13,23} = 0.1 \exp(- i \pi/4)$ and $(V
  \Delta_{Q} V^{\dag})_{ij}$ {for the left-handed down and up
    sectors, respectively}. For the CKM elements,} we use \textsc{CKMfitter}
results \cite{Charles:2015gya}.

{Starting with $\epsilon_K$, we first note that the phase of the SUSY
  contribution to the \kk\ mixing amplitude is essentially twice the
  phase of $\Delta_{Q, 12}$. That is, our choice of $\pi/4$ for this
  phase maximizes the $CP$ phase and is far away from a fine-tuned
  solution to suppress $\epsilon_K$.  We evaluate the MSSM Wilson
  coefficients for $\epsilon_K$ with the $\mathcal{O}(g_s^4, g_s^2 g^2,
  g^4)$ strong and weak contributions \cite{Crivellin:2010ys,
    Altmannshofer:2007cs}. For the RG evolution of the MSSM contribution,
  the LO formula is sufficient \cite{Bagger:1997gg}; lattice results for
  $|\Delta S|=2$ hadronic matrix elements are available from several
  groups \cite{Allton:1998sm}.} 
{For an accurate SM prediction of $\epsilon_K$ one must include all
  NLO corrections \cite{eknlo} and the NNLO contributions involving the
  low charm scale \cite{eknnlo}.  At this level, $\epsilon_K^{\rm SM}$
  agrees with $\epsilon_K^{\rm exp}$, if the value of $|V_{cb}|$
  measured in inclusive $b\to c \ell\nu$ decays is used for the
  calculation of the CKM elements.  Figure~\ref{fig:epsilonK} shows that the
  MSSM can accommodate this situation as well as the scenario with
  $|V_{cb}|$ taken from exclusive $B\to D^{(*)} \ell \nu $ decays
  \cite{Bailey:2014tva}, which calls for a new-physics contribution to
  $\epsilon_K$. The left plot in \fig{fig:epsilonK} clearly reveals that
  the MSSM solution is not fine-tuned but merely requires
  $m_{\tilde{g}}/M_S \gtrsim 1.5$.  For our chosen parameters, we roughly
  find $M_S\gtrsim 3\,\TEV$, with the possibility of slightly lighter
  squarks if the exclusive $|V_{cb}|$ is true.  }

We note that our results are stable if we switch on right-handed squark
mixing as long as $\Delta_{\bar{D}, 12} \lesssim 10^{-5}$.
{Simultaneous sizable left-left and right-right sfermion mixing
  spoils the suppression of gluino box diagrams in
  $\epsilon_K^{\textrm{SUSY}}$ \cite{Crivellin:2010ys}.}  Although in
our scenario $\Delta_{\bar{D}, 12} $ is generated by radiative
corrections, the value is smaller than $10^{-5}$ thanks to the small
down Yukawa coupling.  {A hierarchy $\Delta_{Q,12} \gg
  \Delta_{\bar{D},12}$ appears naturally in UV completions with a
  flavor symmetry; cf.,\ e.g.,\ Refs.~\cite{Hall:1995es, Dudas:1995eq} for models based on the discrete group $S_3$
  and a gauged horizontal U(1), respectively.}

{We next turn to the discussion of $\epsilon_K^\prime$: The thick lines
  in \fig{fig:each_contribution} show the individual contributions to
  $|\epsilon_K^\prime/\epsilon_K|$ for the case of universal squark
  masses. The broken lines show that already a moderate $\bar U$-$\bar
  D$ mass splitting suffices to explain the measured value (indicated by
  the green bands). The second-largest contribution from the
  chromomagnetic penguin diagram comes with a poorly known hadronic
  matrix element~\cite{Bertolini:1994qk}. The $B$ parameter
  parametrizing this matrix element is estimated as $B_G = 1 \pm 3$
  \cite{Buras:1999da}. The yellow band in  \fig{fig:each_contribution}
  is for $1\leq B_G\leq 4$.  Next, we remark that in our parameter region
  the gluino-photon (red line) and chargino-$Z$ (blue line) penguins
  have opposite sign and almost cancel each other. This picture changes
  with nonzero trilinear terms; e.g.,\ $|A_{d,21} |= 0.1 M_S$ ($|
  A_{u,31} A_{u,32} | = 0.1 M^2_S$) can lift the chromomagnetic
  (chargino-$Z$) contribution by about 40\,\% (140\,\%).  We have
  neglected the gluino-$W$ penguin and the gluino-chargino box
  contributions, which matches onto $c_{1,2}^u$ at $\mu=M_S$ and gives
  at most an $\mathcal{O}(10^{-5})$ contribution to
  $\epsilon_K^\prime/\epsilon_K$. }

\begin{figure}[t]
  \begin{center} 
    \includegraphics[width=0.45 \textwidth, 
    bb
     = 0 0 360 362]{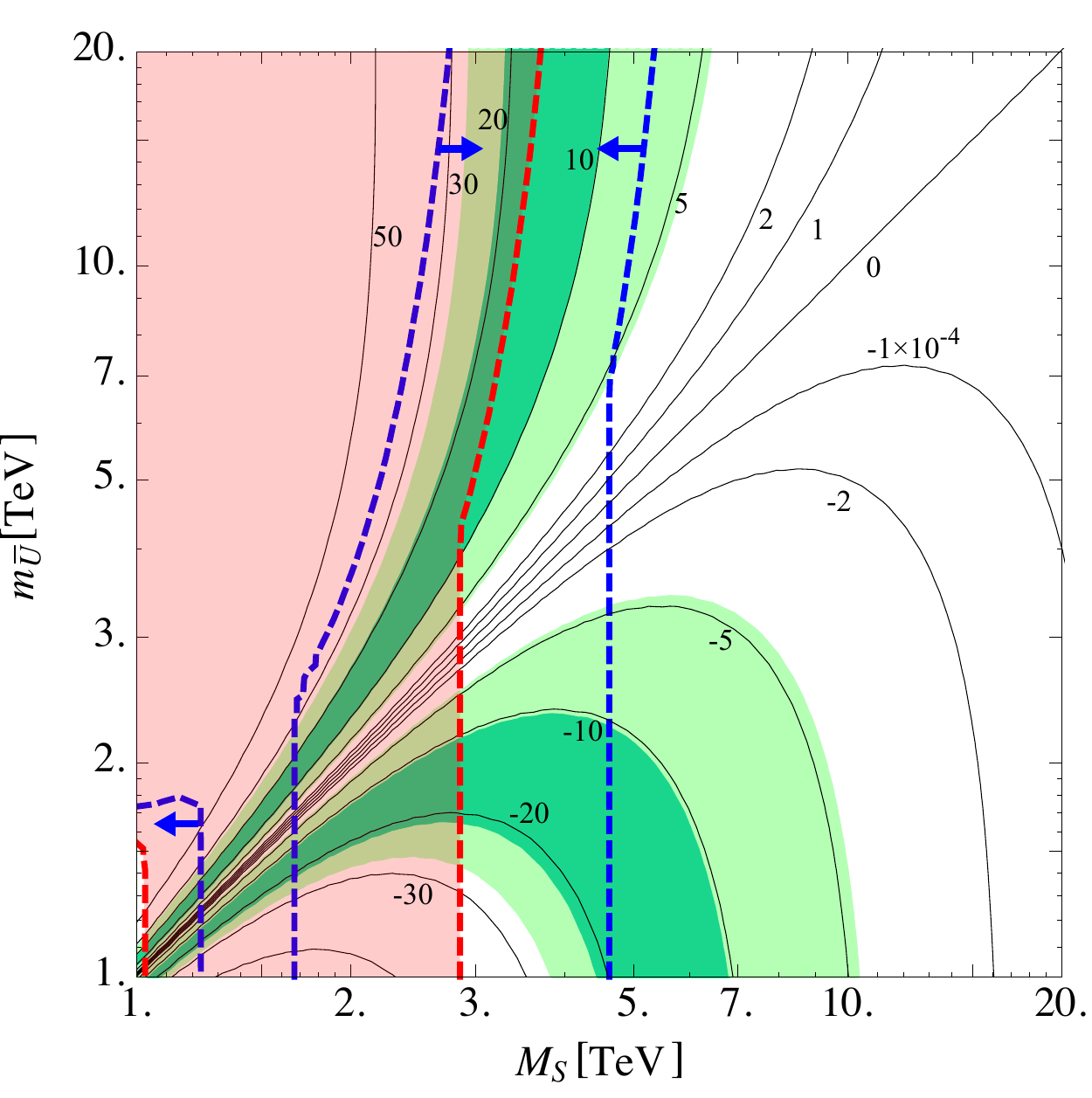}
    \vspace{-0.3cm}
    \caption{Contours of the supersymmetric contributions to
      $\epsilon_K^\prime/\epsilon_K$ in {units} of $10^{-4}$.  The
      $\epsilon_K^\prime/\epsilon_K$ discrepancy {is resolved} at
     $1\,\sigma$\,($2\,\sigma$) in the dark (light) green region.  {The} red shaded
      region is excluded by $\epsilon_K$ with inclusive $|V_{cb}|$ at
      95\,\% C.L., while {the} region between the blue-dashed lines 
      can
      explain the $\epsilon_K$ discrepancy {which is there} for the 
      exclusive $|V_{cb}|$. The green regions labeled with negative
      $\epsilon_K^\prime/\epsilon_K$ correspond to the change 
      $\Delta_{Q,12,13,23}=0.1\exp(-i\pi/4)\to
      \Delta_{Q,12,13,23}=0.1\exp(i3\pi/4)$, which flips the sign of 
       $\epsilon_K^{\prime\rm SUSY}$ (making it positive) while leaving 
     $\epsilon_K$ essentially unchanged.}
\label{fig:contour}
\end{center}
\vspace{-0.8cm}
\end{figure}

Figure~\ref{fig:contour} shows our main result, the portion of the
squark mass plane which simultaneous explains
$\epsilon_K^\prime/\epsilon_K$ and $\epsilon_K$.  The figure uses
  the complete supersymmetric results except for the chromomagnetic
  contribution to $\epsilon^\prime_K$ because of the uncertainty in
  $B_G$.  {The} red region is excluded by {the} measurement of
$\epsilon_K$ at 95\,\% C.L. {in combination with the} inclusive
$V_{cb}$, while {the} region between the blue-dashed lines can explain
the $\epsilon_K$ discrepancy at 95\,\% C.L.\ {for the exclusive value
  of} $|V_{cb}|$.  Note that we also found that there are no constraints
from the mass difference of neutral kaon, $D^0$-$\overline{D}{}^0$
mixing \cite{Golowich:2007ka}, and the neutron electric dipole moment \cite{Baker:2006ts}.

\vspace{-0.4cm}
\section{Conclusions}

In this Letter, {we have calculated $ \epsilon_K^\prime$  
in the MSSM and have shown that the large contributions needed to
solve the discrepancy in \eq{discrepancy} can be obtained for 
squark and gluino masses in the multi-TeV range. The constraint from 
$ \epsilon_K$, which in generic models of new physics precludes large
effects in $ \epsilon_K^\prime$, can be fulfilled without fine-tuning.} 

\vspace{-.4cm}
\section*{Acknowledgments}

We are grateful to {Andrzej Buras}, Motoi Endo, Philipp Frings, Toru
Goto, Satoshi Mishima, {Chris Sachrajda}, and Kei Yamamoto for
fruitful discussions. The work of U.N. is supported by BMBF under Grant
No.~05H15VKKB1. P.T. acknowledges support from the DFG-funded doctoral
school \emph{KSETA}.

\providecommand{\href}[2]{#2}
\begingroup\raggedright

\end{document}